# Modeling of a Cantilever-Based Near-Field Scanning Microwave Microscope

K. Lai, W. Kundhikanjana, M. Kelly, and Z.X. Shen

*Department of Applied Physics and Geballe Laboratory for Advanced Materials, Stanford University, Stanford, CA 94305*

## Abstract

We present a detailed modeling and characterization of our scalable microwave nanoprobe, which is a micro-fabricated cantilever-based scanning microwave probe with separated excitation and sensing electrodes. Using finite-element analysis, the tip-sample interaction is modeled as small impedance changes between the tip electrode and the ground at our working frequencies near 1GHz. The equivalent lumped elements of the cantilever can be determined by transmission line simulation of the matching network, which routes the cantilever signals to 50Ω feed lines. In the microwave electronics, the background common-mode signal is cancelled before the amplifier stage so that high sensitivity (below 1 atto-Farad capacitance changes) is obtained. Experimental characterization of the microwave probes was performed on ion-implanted Si wafers and patterned semiconductor samples. Pure electrical or topographical signals can be realized using different reflection modes of the probe.

# I. Introduction

A new paradigm of electrodynamic measurements, known as near-field microscopy, has emerged in the past few decades to study electromagnetic properties down to a length scale much smaller than the free-space wavelength [1]. Using this technique, sub-micron spatial resolution can be achieved at low frequencies, such as in the microwave regime [1–5]. Here, a sharp near-field probe tip, being an extension of transmission lines or waveguides that carry propagating microwaves, locally interacts with the specimen under test [2]. As the probe scans over the sample surface, variation of the local sample property results in changes of the impedance between the probe tip and ground, which are then detected and recorded to form near-field images, with a spatial resolution comparable to the curvature of the tip apex. Due to the potential applications in electron physics, material science, and biological studies, near-field scanning microwave microscopes have been demonstrated by several groups as scientifically useful instruments [6–11].

The implementation of near-field microwave imaging, however, exhibits several technical challenges. In order to achieve high spatial resolution, the probe has to be brought very close to or in contact with the sample. The fabrication of sharp and robust near-field probes, especially in batch production, is a non-trivial process. Second, the signal from the highly localized tip-sample interaction is usually very small, demanding exceptional sensitivity and stability of the detection system. Thirdly, the near-field results are far more difficult to interpret than the conventional far-field data. The signal here is often a complicated convolution between the probe geometry, especially near the tip end, and the sample property, including the real electromagnetic response and the surface topography [1–5]. Finally, spurious signal can occur due to insufficient shielding of any non-local stray field and the propagating far-field component. As a result, careful design of the tip structure and the detection scheme, as well as extensive characterization of the system, is imperative for the application of near-field microwave imaging.

In a recent publication, we reported the design and preliminary results of a scalable microwave nanoprobe (SMiNa) using micro-machining technique [12]. Based on commercial atomic force microscope (AFM) platforms, the SMiNa resembles the well established scanning capacitance microscope (SCM) technique in that both detection electronics work at a frequency near 1GHz [13, 14]. Our implementation of SMiNa, on the other hand, has several advantages over SCM and other near-field microwave microscopes discussed in the literature. Microwave imaging in general, does not require a back electrode and high quality oxide on the specimen, making possible general-purpose imaging on a variety of samples. The phase information, which is usually not obtained in SCM, is maintained in our microwave electronics, as detailed below. Compared to other cantilever-based microwave microscope designs [15, 16], the SMiNa uses metal lines for the electrodes to greatly reduce the loss in doped Si traces. The background signal is cancelled before amplification to ensure large power gain and high sensitivity. More importantly, our probe is unique in that, besides the sharp sensing electrode, a second electrode surrounding the tip is also present on the cantilever. Various operation modes, including transmission between both electrodes and reflection from either electrode, can be used to achieve different sample information. In this paper, we present a detailed modeling of the contrast mechanism and characterization of the instrument using patterned semiconductor structures. Many experimental findings demonstrated here should lead to interesting applications of the technique.

## II. System characterization

Fig. 1(a) shows a picture of the SMiNa fabricated by Micro-Electro-Mechanical Systems (MEMS) technology. The layer structure of the cantilever was detailed in Ref. [12]. The annular excitation electrode and the tip electrode, which was formed after micro-fabrication by focused-ion beam (FIB) deposition of Pt, can be seen in the scanning electron microscope (SEM) image in Fig. 1(b). Since the dimension of the entire chip is much smaller than the free-space wavelength (30cm) at our working frequency of 1GHz, the lumped-element circuit description is appropriate to represent the probe, with three discrete impedances – $Z_e$ between excitation and ground, $Z_t$ between tip and ground, and $Z_{et}$ between excitation and tip – across the relevant electrodes. Due to the tip-sample interaction, the small impedance changes ($\Delta Z_e$, $\Delta Z_t$, and $\Delta Z_{et}$ in Fig. 1(c)) during the scanning contain local sample information. A complete understanding of the contrast mechanism of our SMiNa, therefore, requires knowledge from three parts of the system. First, the small impedance changes ($\Delta Z_e$, $\Delta Z_t$, and $\Delta Z_{et}$) can be modeled from the tip-sample interaction. Second, both $Z_e$ and $Z_t$ have to be routed to the 50Ω external feed lines so that small variations of them are measurable. Finally, the microwave electronics needs to be sufficiently sensitive and stable to detect small signals due to the impedance changes.

1. Tip-sample interaction

For the current probe design, the aluminum electrodes are not well shielded. The stray field contribution from the exposed electrodes, however, is essentially constant during a single scan and the relative contrast within an image could still be measured in spite of the parasitic effect. For absolute measurements, on the other hand, the signal lines need to be shielded and we are currently fabricating a new set of sensors with extra dielectric and metal layers covering both sides of the cantilever.

Of the three impedance changes, $\Delta Z_t$ represents the most localized tip-sample interaction. Approximating the tip as a conducting sphere with the same diameter at its apex, several groups have performed numerical or analytical analysis to compute this tip-sample impedance change [17–20]. Using empirical calibrations, good agreements with experiments have been obtained, showing the ability of quantitative study of materials in the near-field. However, most existing microwave microscopes utilize chemically etched metal wires as the tips, with a diameter in the order of 10 microns [19 – 22]. For better spatial resolution, our FIB deposited Pt tip has a much smaller diameter 100~200nm, as shown in the inset of Fig. 1(b). Due to the much sharper tip, any meaningful calculation of $\Delta Z_t$ requires extensive knowledge of the exact tip shape and the condition of the sample surface, e. g., the thickness of an unintended contamination layer.

As an illustration of the difficulty involved in quantitative measurements, Fig. 2 shows the simulation results from a FEA software COMSOL3.4 [23] that directly computes the admittance $Y$ (inverse impedance $1/Z$) between two arbitrarily shaped electrodes inside a dielectric or conductive medium. The 2D axisymmetric quasi-static mode and time-harmonic ($f$ = 1GHz) analysis are employed. The tip is assumed to be a perfectly conducting sphere with radius $r$ ranging from 0.1 to 10μm. For demonstration purposes, we calculate the admittance contrast $\Delta Y$ between intrinsic silicon (dielectric constant $\varepsilon_r$ = 12, conductivity ~ 0) and the structure sketched in the inset of Fig. 2, a thin doped Si film (thickness $t$ = 1 μm, $\varepsilon_r$ = 12, resistivity $\rho$ = 1 Ω·cm) on top of the intrinsic substrate. A thin air gap $d$ = 1nm is assumed between the tip and the sample to avoid problems due to finite mesh size. The results, plotted in Fig. 2, show several important features of the near-field tip-sample interaction. As a measure of the capacitive coupling, the imaginary part of the contrast signal roughly scales with the tip diameter, demanding a very sensitive detection scheme when high spatial resolution is required. The real or resistive component of the contrast reduces even faster as a function of the tip size, and is extremely sensitive to the presence of a thin dielectric coating. As a result, for small radii, any slight damage due to tip wear and contamination, as well as indentation on the sample surface, can greatly affect the signal strength. Considering these challenges, we limit ourselves to semi-quantitative comparisons between modeling and experimental data in the following discussions.

When the tip is in contact with the sample, the overall distance between the big excitation electrode and the sample follows the surface topography, resulting in a mostly capacitive $\Delta Z_e$. In other words, the reflection signal from the excitation electrode shows essential the geometrical features, as one would normally obtain from the AFM function. The capacitance change, $\Delta C = \varepsilon_0 (A/h) \Delta h^2$ ~ 20aF, can be easily estimated from the simple parallel-plate capacitor approximation, where $A$ ~ 100 μm$^2$ is the area of the annular electrode, $h$ = 2μm the tip height above it, and a small particle with $\Delta h$ = 100nm on the sample.

Finally, transmission measurements can be performed when the microwave power is supplied by one electrode and collected by the other. Because the coupling impedance $Z_{et}$ is much larger than $Z_e$ or $Z_t$, the power level in the sensing electrode is much lower than that in the excitation electrode, so the common mode signal and shot noise are greatly reduced. In this configuration, all three impedance changes in Fig. 1(c) affect the final signal and one can again use finite element analysis to model $\Delta Z_{et}$ [12]. In this mode, the sample topography will be mixed in due to the height change of the excitation ring with respect to the sample.

## 2. Impedance match

As stated before, the fabricated microwave probe can be modeled as discrete lumped-elements. Neglecting the small geometrical differences between the two electrodes at the tip end, we determined that the overall impedances of the two electrodes and transmission lines on the sensor have a series resistance $R \sim 2\Omega$, inductance $L \sim 2nH$, and a capacitance $C \sim 1pF$ to ground, with the capacitor dominating the total impedance at $f \sim 1GHz$. Because of the large mismatch between $Z_e$ ($Z_t$) and the transmission line impedance $Z_0 = 50\Omega$, microwave power will not be delivered to the probe if it is directly connected to the feed line, and the detection of small impedance variations is not feasible.

Several schemes are available to achieve impedance match in existing microwave microscopes. In our SMiNa, the tip is first attached to the end of a piece of $\lambda/4$ cable to form a transmission line resonator, as seen in Fig. 3(a). The resonator is then critically coupled to the feed line by a parallel open-end tuning stub. Using this tunable stub, nearly perfect match, indicated by a small reflection coefficient S11 < -30dB, is easily obtained, which is important for our common-mode cancellation circuitry, as discussed below. To minimize the force and vibration on the AFM platform, we use very flexible Astro-boa-flex® III microwave cables for both the $\lambda/4$ line and the tuning stub. The nominal group velocity is $2.1 \times 10^8$ m/s in the cable and the loss is 2.4 dB/m.

Fig. 3(b) shows a typical reflection coefficient S11 curve taken by HP 8510B network analyzer. Very good impedance match is achieved at $f = 1.035GHz$ for $l_{\lambda/4} = 4.4cm$ and $l_{stub} = 4.3cm$. The transmission line simulation result using the measured length and propagation constant of the cable is also shown in Fig. 3(b). In fact, $Z_e$ and $Z_t$ can be verified this way by the good fit to standard transmission line formula at various stub length (not shown). Besides, by fitting the S12 data, one can also determine the coupling capacitance $C_{et} \sim 7fF$.

Finally, the output contrast signal from the matching section, which is proportional to ΔS11 (reflection) or ΔS12 (transmission) [24], can be simulated given a small impedance change at the tip. Assuming 1aF capacitive perturbation in either $Z_t$ or $Z_{et}$, we show the calculated ΔS11 and ΔS12 in the inset of Fig. 3(b) and (c). Compared to the situation without the matching section, the signal level is about an order of magnitude higher when $Z_e$ and $Z_t$ are matched to the 50Ω line impedance.

## 3. Microwave circuitry

Fig. 4 shows the block diagram of the detection microwave circuit, which was detailed in Ref. [12]. The reflected or transmitted signal from the matching network contains a large background, which could saturate the amplifiers. In order to sensitively pick up the small contrast from this large background or common-mode signal, we combine a cancellation signal, equal in magnitude but opposite in phase, with the input signal at the amplifier input before a scan. The contrast signal within the scanned area is then amplified by the RF amplifier, demodulated by an IQ mixer, and amplified again in the DC stage. A feedback from the DC output with a long time constant controls the variable attenuators in the cancellation circuit to suppress the system drift due to, e.g., temperature instability. The total gain, including RF and DC, is about 110dB at 1KHz bandwidth. Assuming -20dBm (10μW) power at the tip end, a 1aF capacitance change at the tip produces ~15mV DC output signal, which is above our system noise. Therefore, when properly tuned, our microscope is sensitive to the local impedance change down to the aF level.

The two outputs of the IQ mixer are 90° out-of-phase. For a sample with only dielectric contrast, i.e., a purely imaginary ΔZ, one can adjust the phase of the mixer reference so that the contrast appears only in one channel, the capacitive channel. The other channel, the resistive channel, then corresponds to the real part of ΔZ. Similar to other microwave microscopes, no contrast will show up in the resistive channel if the sample is insulating or a perfect metal [21].

## III. Experimental results

As discussed before, the SMiNa is a very sensitive tool to detect local tip-sample impedance changes down to the sub-micron length scale. Depending on the specific mode, the contrast may originate from the local electrical properties (tip reflection), the sample topography (excitation reflection), or a combination of both (transmission) with much reduced power at the tip electrode. In the following, we show microwave images on several patterned semiconductor samples. The transmission data have been explained in Ref. [12] and results from reflection measurements will be the focus of this study.

1. Reflection from the tip electrode

A set of topography-free samples was fabricated to characterize pure electrical responses from the tip electrode. As shown in Fig. 5(a), $1\times10^{12} cm^{-2}$ phosphorus ions were implanted into repeated 4μm×4μm windows on a high-resistivity (HR, ρ>1000Ω·cm) p-Si substrate. The dose was kept low comparing to other work ($>10^{14} cm^{-2}$) [21, 25, 26] to demonstrate the high sensitivity of the SMiNa. The dopants were then activated at 1100°C for 30 minutes, resulting in a surface doping level ~$2\times10^{16} cm^{-3}$. The flat surface at such low implantation dosage is confirmed by the featureless AFM image, as seen in Fig. 5(b). The two orthogonal microwave images in Fig. 5(c) and (d), on the other hand, show clear contrast between the doped and undoped regions. The contrast mostly shows up in the capacitive channel since the implanted region results in a small shunt resistor that shorts the tip-to-ground capacitance. For the resistive channel image in Fig. 5(d), a line cut shows that the signal is non-monotonic across the boundary, presumably the depletion region, between the p-substrate and the n-type implanted dots.

To compare with the modeling results, identical patterning and implantation process were employed on three p-type Si substrates, labeled as Si-L (ρ = 21Ω·cm, p ~ $6.2\times10^{14} cm^{-3}$), Si-K (ρ = 11Ω·cm, p ~ $1.2\times10^{15} cm^{-3}$), and Si-I (ρ=0.4Ω·cm, p ~ $4.2\times10^{16} cm^{-3}$), where the resistivity was determined by four-probe measurements. The microwave images in the capacitive channel are shown in Fig. 6(a). The contrast between the implanted dots and the substrate follows nicely with the substrate resistivity, and reverses the sign for Si-I, in which the n-type dopants do not fully compensate the background p-doping. In the other output channel (not shown), similar ring-like structure is observed around the dots for Si-L and Si-K, but not Si-I. The simulation result of the total contrast signal (vector sum of the two orthogonal channels) is shown in Fig. 6(b) as a function of the substrate doping density. For simplicity, a uniform doping profile $2\times10^{16}$cm$^{-3}$ for the top 0.5μm is used in the modeling. Good agreement can be obtained between the experimental data and the simulation, in which the following parameters are assumed, tip radius $r$ = 100nm, native oxide thickness $d$ = 1nm, microwave input power $P$ = 10μW, and total gain $G$ = 110dB. To model the tip/sample interaction, we assumed the tip was separated from the sample by a small amount (1nm) to avoid singularities in the finite element computation. For quantitative studies where the sample properties need to be determined *a priori*, calibration of the system using standard kits and careful preparation of the tip and the sample will be needed.

Finally, it is worth pointing out that all images were taken with all light sources turned off. In the presence of significant illumination above the Si band gap, the photo-conductivity of the sample can be so prominent that contrast between the substrate and the implanted region is largely washed out. This interesting photoconductivity effect is currently under intensive study.

2. Reflection from the annular electrode

Fig. 7 shows the reflection images from the large electrode. Several samples was fabricated and measured to demonstrate the dominant dependence on topography of this mode. For the two samples here, the electrical contrast is very different and the tip reflection shows ~0.1V contrast for Al$_2$O$_3$ on Si-HR and ~1.2V for Al on Si-I (not shown). The reflection from the annular electrode, on the other hand, shows the same contrast ~160mV only in the capacitive channel. In fact, very similar contrast of 140 ± 30mV is observed for many test samples with 40 ~ 60nm step height, independent of the sample dielectric constant, conductivity, and the presence of illumination. This contrast signal is also consistent with a simple parallel-plate calculation, which expects ~ 150mV for 50nm step height at the setting of our electronics. We emphasize that, unlike the usual AFM, this mode does not need any optics to measure the sample topography. With sufficient gain and stability, this signal can be utilized to maintain a constant sensor height in the non-contact mode.

**Conclusion**

To summarize, we have developed a procedure to systematically study the contrast mechanism of a cantilever-based scanning microwave probe. The impedance change due to tip-sample interaction and the S-parameters of the matching network can be modeled by finite-element analysis and transmission line simulation, respectively. Our current electronics is optimized for relative imaging rather than absolute measurements. Using a set of ion-implanted Si samples with tip reflection, we have characterized a pure electrical contrast signal that agrees with the modeling. The reflection signal from the annular excitation electrode, consistent with the simulation results, shows essentially the sample topography. Since the tip electrode is most sensitive to local electrical properties and the larger electrode to topography, combining signals from both electrodes can potentially deconvolve the electrical and topographical information, a major problem with earlier microwave instruments.

**Acknowledgement**

The research is funded by the seed grant in Center of Probing the Nanoscale (CPN), Stanford University, with partial support from a gift grant of Agilent Technologies, Inc. and DOE contract DE-FG03-01ER45929-A001. CPN is an NSF NSEC, NSF Grant No. PHY-0425897. The cantilevers and the ion-implantation samples were fabricated in Stanford Nanofabrication Facility (SNF) by A.M. Fitzgerald and B. Chui in A.M. Fitzgerald & Associates, LLC, San Carlos, CA. The authors acknowledge technical support from Agilent Technologies, Inc.

**Reference:**

1. B. T. Rosner and D. W. van der Weide, Rev. Sci. Instrum. **73**, 2505 (2002).
2. M. Fee, S. Chu, T.W. Hansch, Optics Communications 69, 219 (1989).

3. S. M. Anlage, V. V. Talanov, and A. R. Schwartz, in Scanning Probe Microscopy: Electrical and Electromechanical Phenomena at the Nanoscale, edited by S. V. Kalinin and A. Gruverman, Springer, New York, 2006, pp. 207–245.
4. S. M. Anlage, D. E. Steinhauer, B. J. Feenstra, C. P. Vlahacos, and F. C. Wellstood, in Microwave Superconductivity, edited by H. Weinstock and M. Nisenoff Kluwer, Amsterdam, 2001, pp. 239–269.
5. M. Tabib-Azar, Review of Progress in Quantitative Nondestructive Evaluation, 20, 400 (2001).
6. C.P. Vlahacos, R.C. Black, S.M. Anlage, A. Amar, and F.C. Wellstood, Appl. Phys. Lett. **69**, 3272 (1996).
7. M. Tabib-Azar, D.-P. Su, A. Pohar, S.R. LeClair, and G. Ponchak, Rev. Sci. Instrum. **70**, 1725 (1999).
8. M. Golosovsky and D. Davidov, Appl. Phys. Lett. **68**, 1579 (1996).
9. J. Bae, T. Okamoto, T. Fujii, K. Mizuno, and Tatsuo Nozokido, Appl. Phys. Lett. **71**, 3581 (1997).
10. S. Hong, J. Kim, W. Park, and K. Lee, Appl. Phys. Lett. **80**, 524 (2002).
11. T. Wei, X.-D. Xiang, W.G. Wallace-Freedman and P. G. Schultz, Appl. Phys. Lett. **68**, 3506 (1996).
12. K. Lai, M. B. Ji, N. Leindecker, M. A. Kelly, and Z. X. Shen, Rev. Sci. Instrum. **78**, 063702 (2007).
13. R.C. Palmer, E.J. Denlinger, and H. Kawamota, RCA Rev. **43**, 194 (1982).
14. V.V. Zavyalov, J.S. McMurray, and C.C. Williams, Rev. Sci. Instrum. **70**, 158 (1999).
15. D. W. van der Weide, Appl. Phys. Lett. 70, 677 (1997); B.T. Rosner, T. Bork, V. Agrawal, and D.W. van der Weide, Sensors and Actuators A **102**, 185 (2002).
16. M. Tabib-Azar and Y. Wang, IEEE Trans. Microwave Theory Tech. **52**, 971 (2004).
17. A. Imtiaz, M. Pollak, S.M. Anlage, J.D. Barry and J. Melngailis, J. Appl. Phys. **97**, 044302 (2005).
18. C. Gao, B. Hu, P. Zhang, M. Huang, W. Liu, and I. Takeuchi, Appl. Phys. Lett. **84**, 4647 (2004), and references therein.

19. Z. Wang, M.A. Kelly, Z.-X. Shen, G. Wang, X.-D. Xiang, and J.T. Wetzel, J. Appl. Phys. **92**, 808 (2002).
20. J.H. Lee, S. Hyun, and K. Char, Rev. Sci. Instrum. **72**, 1425 (2001).
21. A. Imtiaz and S.M. Anlage, Ultramicroscopy **94**, 209 (2003); *ibid*. J. Appl. Phys. **100**, 44304 (2006).
22. Z. Wang, M.A. Kelly, Z.-X. Shen, L. Shao, W.-K. Chu, and H. Edwards, Appl. Phys. Lett. **86**, 153118 (2005).
23. COMSOL, Inc., Palo Alto, CA.
24. R. Wang, F. Li, and M. Tabib-Azar, Rev. Sci. Instrum. **76**, 054701 (2005), and references therein.
25. A. Tselev, S.M. Anlage, H.M. Christen, R.L. Moreland, V.V. Talanov, and A.R. Schwartz, Rev. Sci. Instrum. **74**, 3167 (2003).
26. A. Imtiaz, S.M. Anlage, J.D. Barry and J. Melngailis, Appl. Phys. Lett.**90**, 143106 (2007).

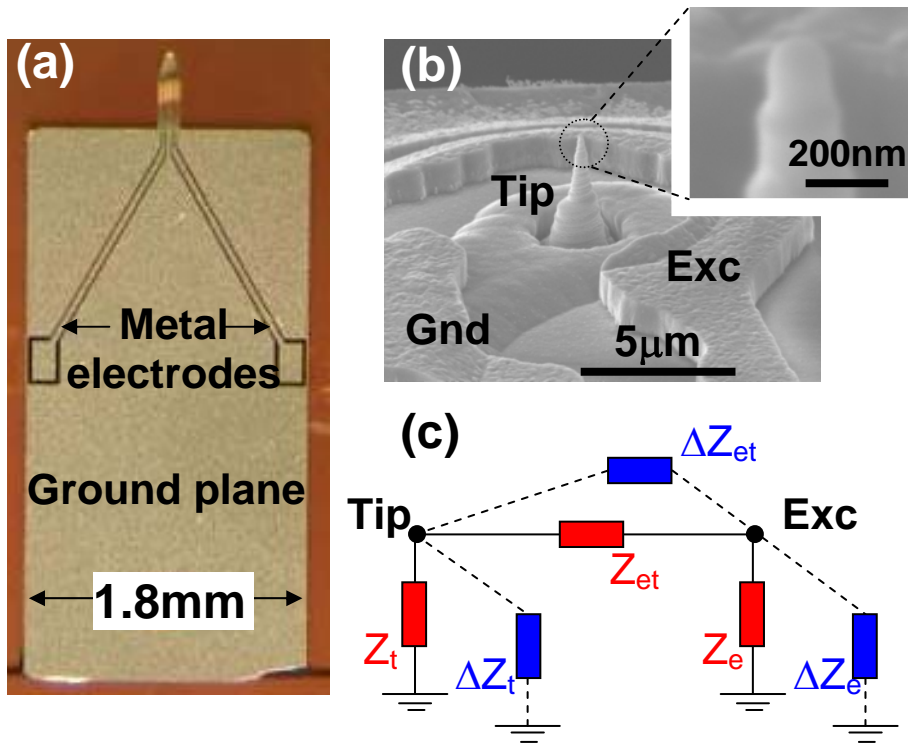

Fig. 1 (a) Micro-fabricated microwave probe with two aluminum electrodes and the ground plane patterned on the Si nitride cantilever. (b) SEM image near the probe end. The apex of the focused-ion beam (FIB) deposited Pt tip is shown in the inset. (c) Lumped-element circuit model of the probe, showing the total and small changes of the impedances.

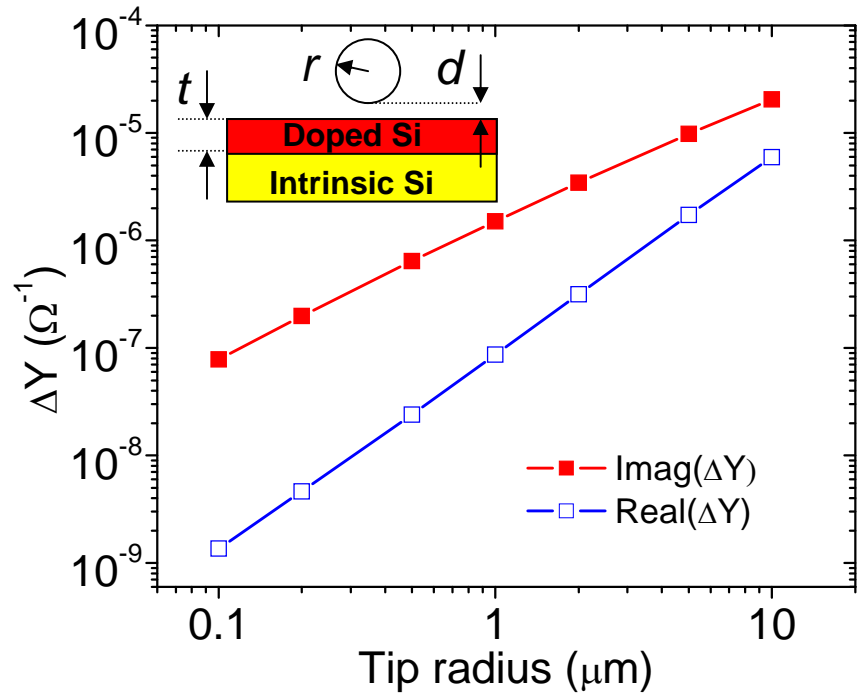

Fig. 2 COMSOL results of the admittance contrast $\Delta Y$ between undoped Si and the structure in the inset, a $t = 1\mu m$ doped ($1\Omega\cdot cm$) Si film on intrinsic substrate. Both real and imaginary parts of $\Delta Y$ are plotted as a function of the tip radius $r$. A very thin air gap $d$ is assumed between the tip and the sample.

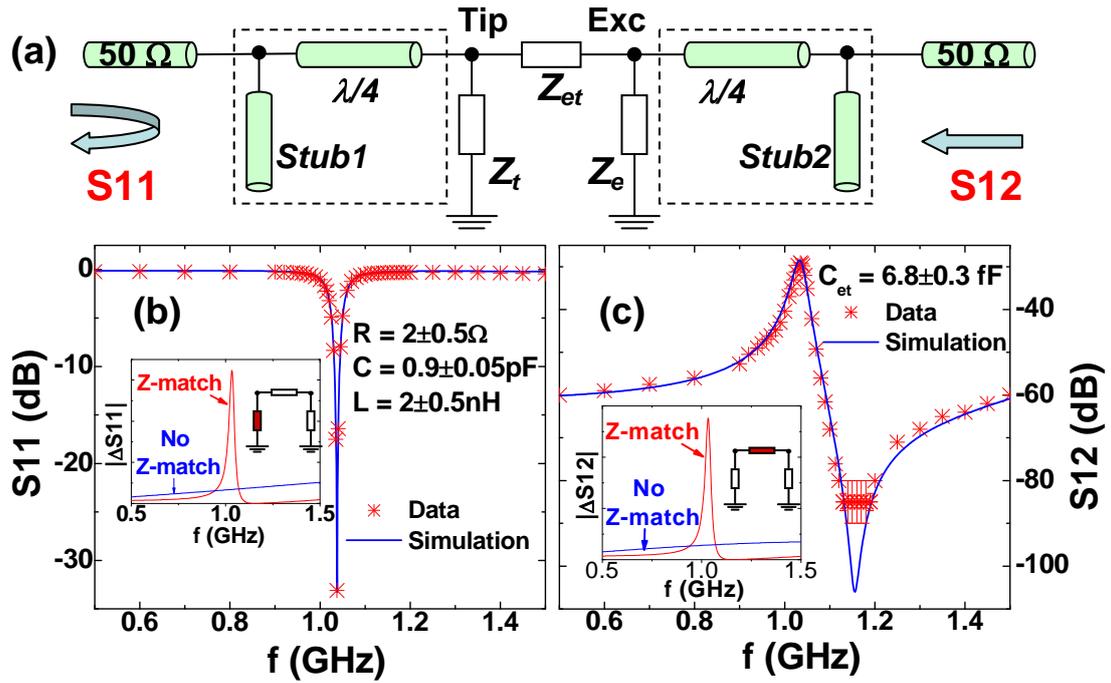

Fig. 3 (a) Impedance match section. A λ/4 cable and a tuning stub (inside the dashed boxes) form the interface between the probe and the 50Ω feed lines. (b) Measured S11 data and a fit to the transmission line analysis. The inset shows |ΔS11| for a given $\Delta C_t$ = 1aF, with and without impedance match. (c) Measured S12 data and the transmission line simulation result, and |ΔS12| for $\Delta C_{et}$ = 1aF in the inset.

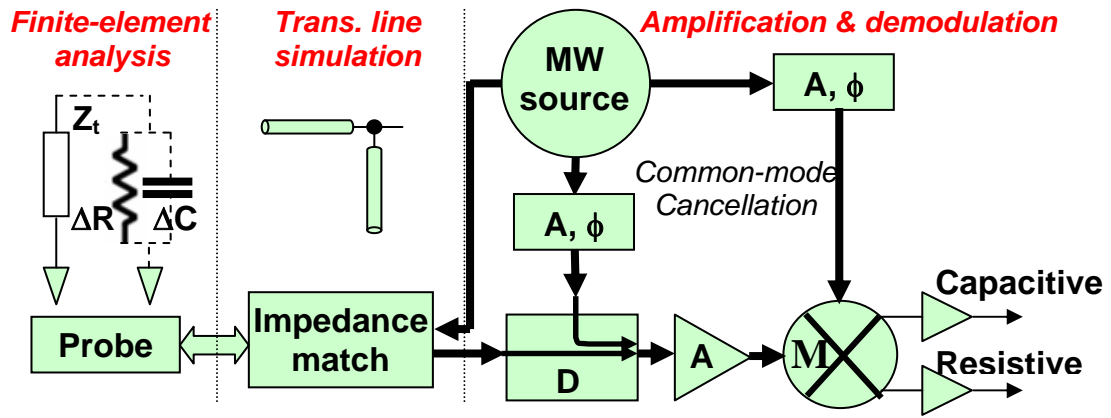

**D** – Directional coupler; **A** – Amplifier; ϕ – Phase shifter; **M** – Mixer

Fig. 4 Block diagram of the microwave detection circuitry. The probe signal due to impedance change, through impedance matching, is null before the scan. The contrast signal is then amplified and demodulated to form near-field images.

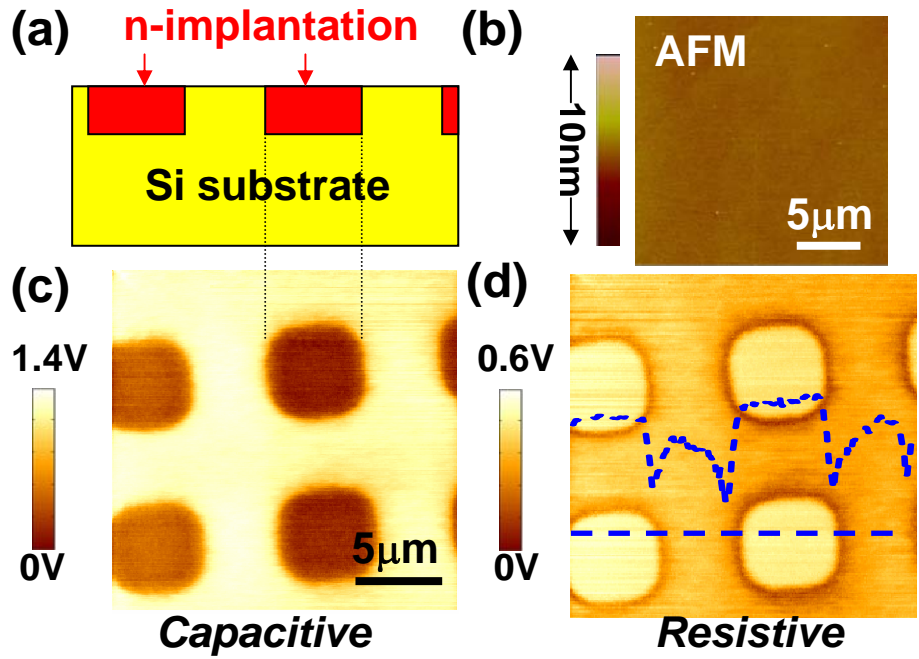

Fig. 5 (a) Schematic of the implantation sample. (b) AFM image of the topography-free sample surface. (c) and (d) Two orthogonal, capacitive and resistive, microwave images (reflection from the tip electrode), with a line cut in the resistive channel image. The color bar corresponds to 1.4V full scale in (c) and 0.6V in (d).

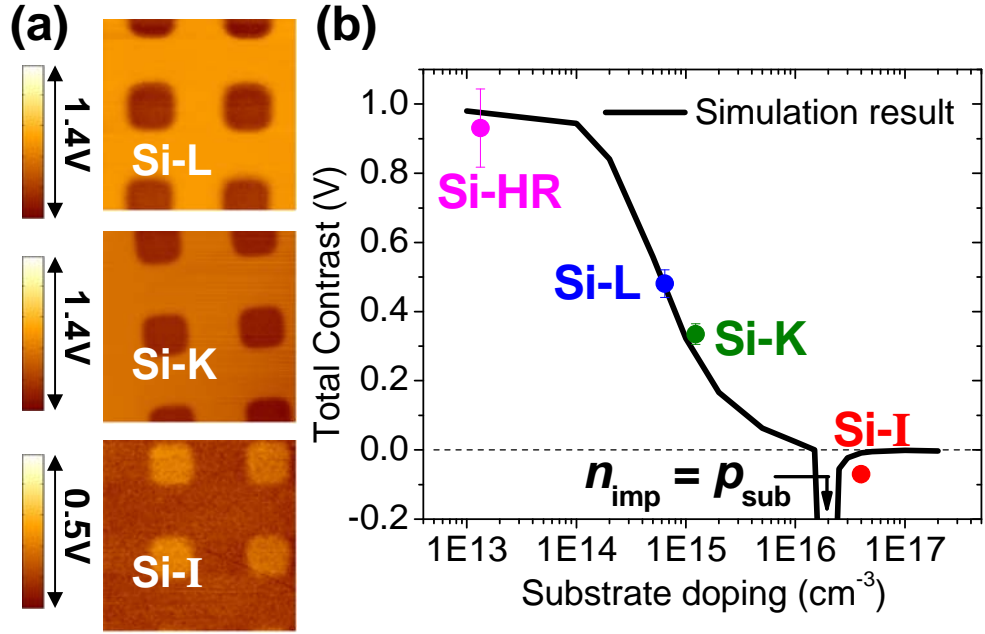

Fig. 6 (a) 20μm × 20μm microwave images (capacitive channel) of three Si samples with the same implantation dose as that in Fig. 5. (b) Simulated total contrast signal between the substrate and the implanted region. The contrast turns negative when the p-type substrate doping level is higher than the surface n-type implantation density. The experimental data are also included for comparison.

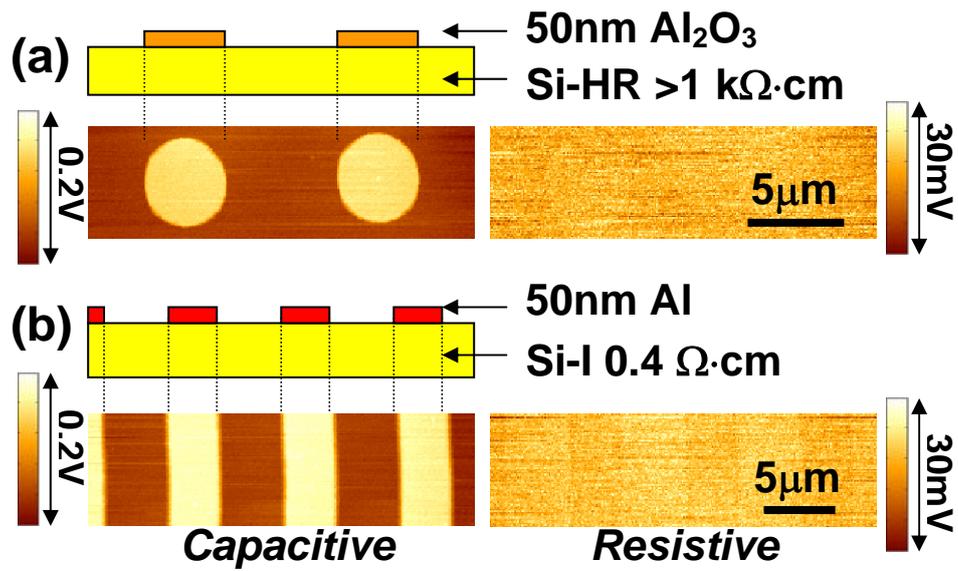

Fig. 7 Microwave images taking reflection signals from the excitation electrode. The sample structures are schematically shown – (a) 50nm $Al_2O_3$ on Si-HR, (b) 50nm Al on Si-I. The full color scale is 200mV for the capacitive channel and 30mV for the resistive channel.

19